\documentclass[twocolumn,superscriptaddress,
showpacs,preprintnumbers,amsmath,amssymb]{revtex4}

\usepackage{amssymb,amsmath,amsthm}
\usepackage{graphicx}
\newtheorem{Thm}{Theorem}
\newtheorem{Cor}{Corollary}
\newtheorem{Lem}{Lemma}
\newtheorem{Prop}{Proposition}
\theoremstyle{definition}
\newtheorem{Exam}{Example}

\newcommand{\bra}[1]{{\left\langle #1 \right|}}
\newcommand{\ket}[1]{{\left| #1 \right\rangle}}

\newcommand{\N}{\mbox{$\mathbb N$}}

\newcommand{\C}{\mbox{$\mathbb C$}}

\newcommand{\x}{\mbox{${\bf x}$}}
\newcommand{\y}{\mbox{${\bf y}$}}
\newcommand{\T}{\mbox{$\mathrm{tr}$}}

\begin{document}
\title{Bound entangled states with nonzero distillable key rate}

\author{Dong~Pyo~Chi}\email{dpchi@math.snu.ac.kr}
\affiliation{
 Department of Mathematical Sciences,
 Seoul National University, Seoul 151-742, Korea
}
\author{Jeong~Woon~Choi}\email{cju@snu.ac.kr}
\affiliation{
 Department of Mathematical Sciences,
 Seoul National University, Seoul 151-742, Korea
}
\author{Jeong~San~Kim}\email{freddie1@snu.ac.kr}
\affiliation{
 Department of Mathematical Sciences,
 Seoul National University, Seoul 151-742, Korea
}
\author{Taewan~Kim}\email{april02@snu.ac.kr}
\affiliation{
 Department of Mathematical Sciences,
 Seoul National University, Seoul 151-742, Korea
}
\author{Soojoon~Lee}\email{level@khu.ac.kr}
\affiliation{
 Department of Mathematics and Research Institute for Basic Sciences,
 Kyung Hee University, Seoul 130-701, Korea
}

\date{\today}

\begin{abstract}
In this paper,
we present sufficient conditions for states to have positive distillable key rate.
Exploiting the conditions, we show that the bound entangled states
given by Horodecki~{\it et al.}~[Phys.~Rev.~Lett. {\bf 94}, 160502 (2005), quant-ph/0506203] have
nonzero distillable key rate,
and finally exhibit new classes of bound entangled states with positive distillable key rate,
but with negative Devetak-Winter lower bound of distillable key rate
for the ccq states of their privacy squeezed versions.
\end{abstract}

\maketitle

\section{Introduction}
Quantum cryptography provides us with a perfectly secure cryptosystem,
which is feasible in a practical way as well as in a theoretical way.
In particular, quantum key distribution among quantum cryptographic protocols
can be considered as one of the most important applications of quantum entanglement,
since secure key distillation in quantum key distribution is
closely related with entanglement distillation~\cite{SP,TKI}.

It has been known that there are two different types of entanglement.
One is called the {\em free} (or, {\em distillable}) entanglement,
from which one can distill a pure entanglement useful for quantum communication
by local quantum operation and classical communication (LOCC),
and the other is called the {\em bound} (or, {\em nondistillable}) entanglement,
which is not distillable.
Even though one cannot distill a pure entanglement useful for quantum communication
from the bound entanglement,
it has been shown that any bound entangled states
can be useful in quantum teleportation~\cite{HHH,Masanes}.
Recently,  Horodecki {\it et al.}~\cite{HHHO1,HHHO2,HPHH}
have shown that
there are some classes of bound entangled states with positive key rate
by showing that the lower bound $K_D^{DW}$ of distillable key rate introduced in~\cite{DW}
is more than zero for those states.

However, the question of whether every entangled state has positive distillable key rate,
$K_D>0$,
has still remained open.
In this paper,
we investigate the properties of quantum states with nonzero distillable key rate, 
and construct several sufficient conditions of $K_D>0$.
Exploiting the conditions,
we show that
the bound entangled states with positive partial transpose (PPT) given in~\cite{HHHO1,HHHO2,HPHH}
have nonzero distillable key rate,
and finally present a new class of PPT bound entangled states satisfying $K_D>0$,
although $K_D^{DW}<0$ for the ccq states of their privacy squeezed versions
presented in~\cite{HHHO2,HPHH}.

This paper is organized as follows.
In Sec.~\ref{Sec:Private States}
we recall the concepts of private states and distillable key rate in~\cite{HHHO2}.
In Sec.~\ref{Sec:Positive_KD}
we construct the sufficient conditions for states with $K_D>0$.
In Sec.~\ref{Sec:Examples}
we show that
the PPT bound entangled states given in~\cite{HHHO1,HHHO2,HPHH}
have nonzero distillable key rate,
and exhibit new classes of PPT bound entangled states satisfying $K_D>0$,
but $K_D^{DW}<0$ for the ccq states of their privacy squeezed versions.
Finally, in Sec.~\ref{Sec:Conclusion} we summarize our results.

\section{Private States and Distillable Key Rate}\label{Sec:Private States}

For a positive integer $d\ge 2$, a {\em private state} (or, {\em pdit}) $\gamma_{ABA'B'}$ on
$\C^{d_A}\otimes \C^{d_{B}}\otimes \C^{d_{A'}}\otimes \C^{d_{B'}}$
with $d_{A}=d_{B}=d$, is defined as
$
\gamma_{ABA'B'} = U \ket{\psi_d^+}\bra{\psi_d^+}\otimes\rho_{A'B'} U^\dagger,
$
where
\begin{equation}
\ket{\psi_d^+}=\frac{1}{\sqrt{d}}\sum_{k=0}^{d-1} \ket{k k}_{AB},
\label{eq:max_ent_state}
\end{equation}
$\rho_{A'B'}$ is an arbitrary state of the subsystem $A'B'$,
and $U$ is an arbitrary twisting operation
\begin{equation}
U=\sum_{k,l=0}^{d-1} \ket{k l}\bra{k l}\otimes U_{kl}
\label{eq:twisting}
\end{equation}
with unitary matrices $U_{kl}$.
Then $\gamma_{ABA'B'}$ can be rewritten as
\begin{equation}
\gamma_{ABA'B'} = \frac{1}{d} \sum_{k,l=0}^{d-1} \ket{k k}\bra{ll}\otimes U_{kk}\rho_{A'B'} U^{\dagger}_{ll}.
\label{eq:pdit}
\end{equation}
When $d=2$, $\gamma_{ABA'B'}$ is called a {\em private bit} (or, {\em pbit}).
Then we can have the following proposition~\cite{HHHO2}.

\begin{Prop}\label{Prop:equiv2}
If a state $\rho \in \mathcal{B}(\C^2 \otimes \C^2 \otimes \C^d \otimes \C^{d'})$
with $\rho=\sum_{i,j,k,l}\ket{ij}\bra{kl} \otimes A_{ijkl}$ fulfills
$\|A_{0011}\|\ge 1/2-\varepsilon$,
then for $0<\varepsilon<1$ there exists a pbit $\gamma$ such that
$\|\rho-\gamma\| \le \delta(\varepsilon)$
with $\delta(\varepsilon)$ vanishing, when $\varepsilon$ approaches zero.
More specifically,
\begin{equation}
\delta(\varepsilon)=\sqrt{\ln{2\left(8\sqrt{2\varepsilon}+h(2\sqrt{2\varepsilon})\right)}}
+2\sqrt{2\varepsilon},
\label{eq:delta}
\end{equation}
where $h$ is the binary entropy function.
\end{Prop}

Now, we define the {\em distillable key rate} $K_D$ as presented in~\cite{HHHO2}.
Let $\rho_{AB}$ be a given state in ${\cal B}(\C^{d_A}\otimes {\C}^{d_B})$.
For each positive integer $n$, consider a sequence $P_n$ of
LOCC operations such that $P_n(\rho_{AB}^{\otimes n})$ is a state
in ${\cal B}(\C^{d_n}\otimes {\C}^{d_n})$.
The family of the operations ${\cal P} \equiv \left\{P_n: n\in \N\right\}$ is called
a {\em pdit distillation protocol} of $\rho_{AB}$ if 
\begin{equation}
\lim_{n\rightarrow \infty} \|P_n(\rho_{AB}^{\otimes n})-\gamma_{d_n}\| = 0,
\label{eq:protocol00}
\end{equation}
where $\gamma_{d_n}$ is a pdit whose $AB$ part is of dimension $d_n^2$.
The {\em rate of a protocol} $\cal P$ is given by
\begin{equation}
R_{\cal P}=\limsup_{n\rightarrow \infty} {\log d_n \over n},
\label{eq:rate_protocol}
\end{equation}
and the {\em distillable key rate} of $\rho_{AB}$ is defined as
the maximum rate of a protocol
\begin{equation}
K_D(\rho_{AB})=\sup_{\cal P}R_{\cal P}.
\label{eq:d_key_rate}
\end{equation}
Then the following proposition for the distillable key rate $K_D$ can be obtained,
as shown in~\cite{HHHO2}.
\begin{Prop}\label{Prop:Krate00}
If a state $\rho$ is close enough to a pbit in trace norm, then $K_D (\rho)>0$.
\end{Prop}
%

\section{Sufficient Conditions for Positive Distillable Key Rate}\label{Sec:Positive_KD}
Proposition~\ref{Prop:Krate00} in Sec.~\ref{Sec:Private States}
provides us with a simple sufficient condition of $K_D (\rho)>0$, as follows.

\begin{Lem}\label{Lem:Krate01}
If one can transform, by LOCC, such as the recurrence protocol,
sufficiently many copies of a state $\rho$ into a state close enough
to a private state in trace norm, then $K_D (\rho)>0$.
\end{Lem}
\begin{proof}
Assume that, by LOCC, the state of sufficiently many copies of $\rho$
is transformed into $\rho'$, which is close enough to a pdit in trace norm.
Then, by Proposition~\ref{Prop:Krate00},
$\rho'$ has a nonzero distillable key rate, that is, $K_D(\rho')>0$.
Thus, by the definition of $K_D$,
there exists a family of LOCC operations $\mathcal{P}_0$ such that
\begin{equation}
{R}_{\mathcal{P}_0}=\limsup_{n\rightarrow \infty} \frac{\log d_n}{n}
\label{eq:Krate01}
\end{equation}
is nonzero for $\rho'$.

Now, let us consider $K_D(\rho)$ in accordance with $K_D(\rho')$.
By the assumption, $\rho'$ can be made out of
sufficiently many $m$ copies of $\rho$ by LOCC,
which is denoted  by $P'$.
We let $\mathcal{P}=\{P\circ P': P\in \mathcal{P}_0\}$.
Then $\mathcal{P}$ is a pdit distillation protocol of $\rho$,
and hence we clearly obtain
\begin{equation}
K_D(\rho)\ge {R}_{\mathcal{P}}=\limsup_{n\rightarrow \infty} \frac{\log d_n}{mn}
=\frac{{R}_{\mathcal{P}_0}}{m}>0.
\label{eq:Krate_main}
\end{equation}
Therefore, this completes the proof.
\end{proof}

By Lemma~\ref{Lem:Krate01},
we have an explicit form of a sufficient condition for $K_D>0$.
\begin{Thm}\label{Thm:RP}
Let $\rho$ be any state in $\mathcal{B}(\C^2 \otimes \C^2 \otimes \C^d \otimes \C^{d})$
with $\rho=\sum_{i,j,k,l=0}^{1}\ket{ij}\bra{kl} \otimes A_{ijkl}$.
If $\|A_{0000}\|=\|A_{0011}\|=\|A_{1111}\|$ and $\|A_{0101}\|<\|A_{0011}\|$, $\|A_{1010}\|<\|A_{0011}\|$,
then $K_D(\rho)>0$.
\end{Thm}

\begin{proof}
Repeating the recurrence protocol on many copies of $\rho$,
by Lemma~\ref{Lem:RP} in Appendix, we can obtain
\begin{equation}
\rho'
=\frac{1}{N}
\left[%
\begin{array}{cccc}
  A_{0000}^{\otimes n} & A_{0001}^{\otimes n} & A_{0010}^{\otimes n} & A_{0011}^{\otimes n} \\
  A_{0100}^{\otimes n} & A_{0101}^{\otimes n} & A_{0110}^{\otimes n} & A_{0111}^{\otimes n} \\
  A_{1000}^{\otimes n} & A_{1001}^{\otimes n} & A_{1010}^{\otimes n} & A_{1011}^{\otimes n} \\
  A_{1100}^{\otimes n} & A_{1101}^{\otimes n} & A_{1110}^{\otimes n} & A_{1111}^{\otimes n}
\end{array}%
\right],
\label{eq:RP00}
\end{equation}
where $N=\|A_{0000}\|^n+\|A_{0101}\|^n+\|A_{1010}\|^n+\|A_{1111}\|^n$.
Then we have
$\|A'_{0011}\|=\|A_{0011}\|^n/N$,
where $A'_{0011}$ is the upper-right block of $\rho'$.

Since $\|A_{0000}\|=\|A_{0011}\|=\|A_{1111}\|$,
$\|A_{0101}\|<\|A_{0011}\|$, and $\|A_{1010}\|<\|A_{0011}\|$,
we can readily show that
$\|A'_{0011}\|$ converges to $1/2$ as $n$ tends to infinity.
Therefore, by Proposition~\ref{Prop:equiv2} and Lemma~\ref{Lem:Krate01},
we conclude that $K_D(\rho)$ is positive.
\end{proof}

By Theorem~\ref{Thm:RP},
we clearly obtain the following corollary.
\begin{Cor}\label{Cor:RP}
Let $\rho$ be a state in
$\mathcal{B}(\C^2 \otimes \C^2 \otimes \C^d \otimes \C^{d})$
of the form
\begin{eqnarray}
\rho&=&\ket{\phi^+}\bra{\phi^+} \otimes \sigma_0
+\ket{\phi^-}\bra{\phi^-} \otimes \sigma_1\nonumber\\
&&+\ket{\psi^+}\bra{\psi^+} \otimes \sigma_2
+\ket{\psi^-}\bra{\psi^-} \otimes \sigma_3,
\label{eq:Belldiag_rho}
\end{eqnarray}
where  $\ket{\phi^{\pm}}$ and $\ket{\psi^{\pm}}$ are Bell states in $\C^2\otimes \C^2$.
Then if $\|\sigma_0-\sigma_1\|>1/2$ and $\T(\sigma_0\sigma_1)=0$,
then $K_{D}(\rho)>0$.
\end{Cor}
\begin{proof}
$\rho$ has the following matrix form:
\begin{equation}
\rho
=\frac{1}{2}\left[%
\begin{array}{cccc}
  \sigma_0+\sigma_1 & 0 & 0 & \sigma_0-\sigma_1 \\
  0 & \sigma_2+\sigma_3 & \sigma_2-\sigma_3 & 0 \\
  0 & \sigma_2-\sigma_3 & \sigma_2+\sigma_3 & 0 \\
  \sigma_0-\sigma_1 & 0 & 0 & \sigma_0+\sigma_1 \\
\end{array}%
\right].
\label{eq:Belldiag_rho2}
\end{equation}
By Lemma~\ref{Lem:orthogonal} in Appendix,
we have $\|\sigma_0-\sigma_1\|=\|\sigma_0+\sigma_1\|$,
and hence $\|\sigma_2+\sigma_3\|<1/2<\|\sigma_0+\sigma_1\|$.
Therefore, since all the hypotheses in Theorem~\ref{Thm:RP} are satisfied,
we conclude that $K_D(\rho)>0$.
\end{proof}

Now, let us consider the {\em privacy squeezed} state $\sigma_{AB}$ of $\rho$ in Eq.~(\ref{eq:Belldiag_rho}),
which has been introduced in~\cite{HHHO2,HPHH}, is
\begin{equation}
\sigma_{AB}
=\frac{1}{2}
\left[%
\begin{array}{cccc}
  \|\sigma_0+\sigma_1\| & 0 & 0 & \|\sigma_0-\sigma_1\| \\
  0 & \|\sigma_2+\sigma_3\| & \|\sigma_2-\sigma_3\| & 0 \\
  0 & \|\sigma_2-\sigma_3\| & \|\sigma_2+\sigma_3\| & 0 \\
  \|\sigma_0-\sigma_1\| & 0 & 0 & \|\sigma_0+\sigma_1\|
\end{array}%
\right],
\label{eq:p_squeezed11}
\end{equation}
and let $\ket{\Psi}_{ABE}$ be a purification of $\sigma_{AB}$.
Then
\begin{eqnarray}
\ket{\Psi}_{ABE}
&=&\sqrt{x}\ket{\phi^+}\ket{e_0}
+\sqrt{y}\ket{\phi^-}\ket{e_1}\nonumber \\
&&+\sqrt{z}\ket{\psi^+}\ket{e_2}+\sqrt{w}\ket{\psi^-}\ket{e_3}\nonumber \\
&=&\frac{1}{2}\ket{00}\otimes\left(\sqrt{x}\ket{e_0}+\sqrt{y}\ket{e_1}\right)\nonumber \\
&&+\frac{1}{2}\ket{11}\otimes\left(\sqrt{x}\ket{e_0}-\sqrt{y}\ket{e_1}\right)\nonumber \\
&&+\frac{1}{2}\ket{01}\otimes\left(\sqrt{z}\ket{e_2}+\sqrt{w}\ket{e_3}\right)\nonumber \\
&&+\frac{1}{2}\ket{10}\otimes\left(\sqrt{z}\ket{e_2}-\sqrt{w}\ket{e_3}\right),
\end{eqnarray}
where
\begin{eqnarray}
x&=&\frac{1}{2}\left(\|\sigma_0+\sigma_1\|+\|\sigma_0-\sigma_1\|\right),\nonumber\\
y&=&\frac{1}{2}\left(\|\sigma_0+\sigma_1\|-\|\sigma_0-\sigma_1\|\right),\nonumber\\
z&=&\frac{1}{2}\left(\|\sigma_2+\sigma_3\|+\|\sigma_2-\sigma_3\|\right), \nonumber\\
w&=&\frac{1}{2}\left(\|\sigma_2+\sigma_3\|-\|\sigma_2-\sigma_3\|\right).
\label{eq:abcd}
\end{eqnarray}
By simple calculations,
we can know that the ccq state $\sigma^{ccq}_{ABE}$ of $\ket{\Psi}_{ABE}$ is
\begin{equation}
\sigma^{ccq}_{ABE}=\frac{1}{2}\sum_{i,j=0}^{1}\ket{ij}\bra{ij}\otimes P_{ij},
\label{eq:ccq}
\end{equation}
where $P_{00}$, $P_{11}$, $P_{01}$, and $P_{10}$ are the projections onto the subspaces spanned by
$\sqrt{x}\ket{e_0}+\sqrt{y}\ket{e_1}$,
$\sqrt{x}\ket{e_0}-\sqrt{y}\ket{e_1}$,
$\sqrt{z}\ket{e_2}+\sqrt{w}\ket{e_3}$, and
$\sqrt{z}\ket{e_2}-\sqrt{w}\ket{e_3}$, respectively.

We note that one can get
$K_D^{DW}=I(A:B)-I(A:E)$ bits of key
for the ccq state obtained from the state $\ket{\Psi}_{ABE}$
by Devetak-Winter~\cite{DW} protocol,
where $I(A:B)=S(A)+S(B)-S(AB)$, $S$ being von Neumann entropy.
Therefore, by straightforward calculations, one can obtain
\begin{equation}
K_D^{DW}(\sigma_{ABE}^{ccq})
=1-S(E),
\label{eq:SE0}
\end{equation}
and
\begin{equation}
S(E)=-x\log_2x-y\log_2y-z\log_2z-w\log_2w.
\label{eq:SEabcd1}
\end{equation}
Hence, we obtain the following lemma.

\begin{Lem}\label{Lem:SE}
Let $\rho$ be a state in
$\mathcal{B}(\C^2 \otimes \C^2 \otimes \C^d \otimes \C^{d})$
of the form in {\em Eq.~(\ref{eq:Belldiag_rho})},
and let $\sigma^{ccq}_{ABE}$ be the ccq state
obtained from the privacy squeezed state of $\rho$.
Then $K_{D}^{DW}(\sigma_{ABE}^{ccq})= 1-S(E)$,
and furthermore,
\begin{equation}
S(E)=-x\log_2x-y\log_2y-z\log_2z-w\log_2w,
\label{eq:SEabcd}
\end{equation}
where $x$, $y$, $z$, and $w$ are in {\em Eq.~(\ref{eq:abcd})}.
\end{Lem}

We now present another sufficient condition of $K_D>0$,
which is a generalization of a result of Horodecki {\it et al.}
(Proposition~1~in~\cite{HPHH}).
\begin{Thm}\label{Thm:KDW}
Let $\rho$ be any state in $\mathcal{B}(\C^2 \otimes \C^2 \otimes \C^d \otimes \C^{d})$
with $\rho=\sum_{i,j,k,l=0}^{1}\ket{ij}\bra{kl} \otimes A_{ijkl}$,
and let
\begin{eqnarray}
x&=&(\|A_{0000}\|+\|A_{1111}\|)/2+\|A_{0011}\|,\nonumber\\
y&=&(\|A_{0000}\|+\|A_{1111}\|)/2-\|A_{0011}\|,\nonumber\\
z&=&(\|A_{0101}\|+\|A_{1010}\|)/2+\|A_{0110}\|,\nonumber\\
w&=&(\|A_{0101}\|+\|A_{1010}\|)/2-\|A_{0110}\|.
\label{eq:KDW00}
\end{eqnarray}
If $-x\log_2x-y\log_2y-z\log_2z-w\log_2w < 1$, then $K_D(\rho)>0$.
More specifically,
\begin{eqnarray}
K_D(\rho) \ge
1&+&x\log_2x+y\log_2y\nonumber\\
&+&z\log_2z+w\log_2w > 0.
\label{eq:KDW01}
\end{eqnarray}
\end{Thm}
\begin{proof}
$\rho$ has the matrix form
\begin{equation}
\rho = \left[%
\begin{array}{cccc}
  A_{0000} & A_{0001} & A_{0010} & A_{0011} \\
  A_{0100} & A_{0101} & A_{0110} & A_{0111} \\
  A_{1000} & A_{1001} & A_{1010} & A_{1011} \\
  A_{1100} & A_{1101} & A_{1110} & A_{1111} \\
\end{array}%
\right].
\label{eq:KDW01}
\end{equation}
If we apply an appropriate twisting operation first, then we
can get
\begin{equation}
\rho_{tw}
=\left[%
\begin{array}{cccc}
  B_{0000} & B_{0001} & B_{0010} & B_{0011} \\
  B_{0100} & B_{0101} & B_{0110} & B_{0111} \\
  B_{1000} & B_{1001} & B_{1010} & B_{1011} \\
  B_{1100} & B_{1101} & B_{1110} & B_{1111}
\end{array}%
\right],
\label{eq:rhotw}
\end{equation}
 where $B_{0000}$, $B_{1111}$, $B_{0011}$, $B_{1100}$,
$B_{0101}$, $B_{1010}$, $B_{0110}$, and $B_{1001}$ are positive and
\begin{eqnarray}
\T{B_{0000}}=\|A_{0000}\|,&& \T{B_{1111}}=\|A_{1111}\|,\nonumber\\
\T{B_{0011}}=\|A_{0011}\|,&& \T{B_{1100}}=\|A_{1100}\|,\nonumber\\
\T{B_{0101}}=\|A_{0101}\|,&& \T{B_{1010}}=\|A_{1010}\|,\nonumber\\
\T{B_{0110}}=\|A_{0110}\|,&& \T{B_{1001}}=\|A_{1001}\|.
\label{eq:twisting}
\end{eqnarray}
By the same LOCC on the subsystem $AB$ as the depolarization in $\C^{2}\otimes \C^{2}$,
we can get the following state
\begin{equation}
\tilde{\rho}_{tw} = \left[%
\begin{array}{cccc}
  \frac{B_{0000}+B_{1111}}{2}& 0 & 0 & \frac{B_{0011}+B_{1100}}{2} \\
  0 & \frac{B_{0101}+B_{1010}}{2} & \frac{B_{0110}+ B_{1001}}{2}&0 \\
  0 & \frac{B_{0110}+B_{1001}}{2} & \frac{B_{0101}+ B_{1010}}{2}&0 \\
  \frac{B_{0011}+B_{1100}}{2}& 0 & 0 & \frac{B_{0000}+ B_{1111}}{2}\\
\end{array}%
\right].
\label{eq:tilde_rho}
\end{equation}
Let $\sigma^{tw}_{AB}$ be the privacy squeezed state of $\tilde{\rho}_{tw}$,
and $\sigma^{ccq}_{ABE}$ be the ccq state obtained from $\sigma^{tw}_{AB}$.
We remark that
the distillable key rate of $\rho_{tw}$
is the same as that of the original state $\rho$,
and furthermore
$\sigma^{ccq}_{ABE}$ has the key rate no better than
that of $\rho_{tw}$~\cite{HHHO1,HHHO2,HPHH}.
Since $\tilde{\rho}_{tw}$ is of the form in Eq.~(\ref{eq:Belldiag_rho}),
for $x$, $y$, $z$, and $w$ in Eq.~(\ref{eq:KDW00}),
we straightforwardly obtain
\begin{eqnarray}
K_D^{DW}(\sigma^{ccq}_{ABE})&=&1-S(E)\nonumber\\
&=&1+x\log_2x+y\log_2y\nonumber\\
&&+z\log_2z+w\log_2w,
\label{eq:K_DW}
\end{eqnarray}
by Lemma~\ref{Lem:SE}.
Since $-x\log_2x-y\log_2y-z\log_2z-w\log_2w < 1$ by our hypothesis,
we have
\begin{equation}
K_D(\rho) \ge K_D^{DW}(\sigma^{ccq}_{ABE})
= 1-S(E)> 0.
\label{eq:KDW_final}
\end{equation}
\end{proof}

\section{Examples}\label{Sec:Examples}

We first consider the PPT states with $K_D>0$ presented in~\cite{HHHO1,HHHO2}.
\begin{Exam}\label{Ex:1}
Let $\varrho_s=2P_{sym}/({d^2+d})$ and
$\varrho_a=2P_{as}/({d^2-d})$ with the symmetric projector $P_{sym}$
and the antisymmetric projector $P_{as}$ on $\C^{d}\otimes \C^{d}$, and
\begin{equation}
\rho=\frac{1}{2}
\left[%
\begin{array}{cccc}
  p({\tau_1+\tau_0}) &0&0&p({\tau_1-\tau_0}) \\
0& (1-2p)\tau_0&0&0 \\
0&0&(1-2p)\tau_0& 0\\
p{({\tau_1-\tau_0})} &0&0& p({\tau_1+\tau_0})  \\
\end{array}%
\right],
\label{eq:ex1}
\end{equation}
where $\tau_0=\varrho_s^{\otimes l}$ and
$\tau_1=[(\varrho_a+\varrho_s)/2]^{\otimes l}$.
Then we can obtain
\begin{widetext}
\begin{equation}
\rho'=\dfrac{1}{2^mN}
\left[%
\begin{array}{cccc}
   \left[p({\tau_1+\tau_0})\right]^{\otimes m} &0&0& \left[p({\tau_1-\tau_0})\right]^{\otimes m} \\
0& \left[(1-2p)\tau_0\right]^{\otimes m}&0&0 \\
0&0&\left[(1-2p)\tau_0\right]^{\otimes m}& 0\\
\left[p{({\tau_1-\tau_0})}\right]^{\otimes m} &0&0& \left[p({\tau_1+\tau_0})\right]^{\otimes m}  \\
\end{array}%
\right],
\label{eq:ex11}
\end{equation}
\end{widetext}
with $N=2p^m+2(1/2-p)^m$, from $\rho$ by the recurrence protocol.
Then it follows from Lemma~\ref{Lem:PPT} in Appendix that
the state $\rho$ is PPT
for $p\in [0,1/3]$ and $(1-p)/p \geq \left[d/(d-1)\right]^l$,
and hence the state $\rho'$ is also PPT.

Let
\begin{eqnarray}
x&=&\frac{1}{2^mN}\left[\left\|p(\tau_1+\tau_0)\right\|^m
+\left\|p(\tau_1-\tau_0)\right\|^m\right]
\nonumber\\
y&=&\frac{1}{2^mN}\left[\left\|p(\tau_1+\tau_0)\right\|^m
-\left\|p(\tau_1-\tau_0)\right\|^m\right]
\nonumber\\
z&=&w=\frac{1}{2^mN}\left\|(1-2p)\tau_0\right\|^m.
\label{eq:xyzw_ex1}
\end{eqnarray}
Then for $p \in (1/4,1/3]$, by choosing sufficiently large $m$ and $l$,
we have
$-x\log_2x-y\log_2y-z\log_2z-w\log_2w <1$.
Therefore, we can obtain the PPT states with $K_D>0$ by Theorem~\ref{Thm:KDW}.
\end{Exam}

We consider the low-dimensional PPT states with $K_D>0$ presented in~\cite{HPHH}.
\begin{Exam}
For two private bits $\gamma_1$ and $\gamma_2$,
take any biased mixture of the form:
\begin{equation}
\rho=p_1\gamma_1+p_2\sigma_x^A\gamma_2\sigma_x^A
\label{eq:low_dim}
\end{equation}
with $p_1>p_2$ and $\sigma_x^A=[\sigma_x]_A \otimes I_{A'BB'}$,
where $\sigma_x$ is one of Pauli matrices representing the bit flip.
Then $\rho$ has the following matrix form.
\begin{equation}
\rho=\frac{1}{2}\left[%
\begin{array}{cccc}
  p_1\sqrt{X_1X_1^{\dagger}} & 0 & 0 & p_1X_1 \\
  0 & p_2\sqrt{X_2X_2^{\dagger}} & p_2X_2 & 0\\
  0 & p_2X_2^{\dagger} & p_2\sqrt{X_2^{\dagger}X_2} & 0 \\
  p_1X_1^{\dagger} & 0 & 0 & p_1\sqrt{X_1^{\dagger}X_1} \\
\end{array}%
\right],
\label{eq:low_dim1}
\end{equation}
where $X_1$ and $X_2$ are arbitrary operators with trace norm one.
Then we can easily show that
the values $x$, $y$, $z$, and $w$ in Theorem~\ref{Thm:KDW}
are $x=p_1$, $y=0$, $z=p_2$, and $w=0$.
By Theorem~\ref{Thm:KDW}, $K_D(\rho)>0$ since $p_1+p_2=1$ and $p_1>p_2$.
\end{Exam}

We present the PPT states with $K_D>0$ which can be shown by Theorem~\ref{Thm:RP}.
\begin{Exam}\label{Ex:300}
For $0<q<{(2-\sqrt{2})}/{8}$, let
\begin{equation}
p=\frac{1-2q}{4+2\sqrt{2}},
\label{eq:ex300}
\end{equation}
\begin{eqnarray}
\sigma_0&=&p\left(\ket{\phi^+}\bra{\phi^+}+\ket{01}\bra{01}\right), \nonumber\\
\sigma_1&=&p\left(\ket{\phi^-}\bra{\phi^-}+\ket{10}\bra{10}\right),
\label{eq:ex30}
\end{eqnarray}
and let $\Gamma$ denote partial transposition over the subsystem $BB'$.
Then we have $\T(\sigma_0\sigma_1)=0$,
\begin{eqnarray}
\sigma_0+\sigma_1 &=& p\mathcal{I} = \sigma_0^\Gamma + \sigma_1^\Gamma, \nonumber\\
\sigma_0-\sigma_1 &=&
p\left(\ket{00}\bra{11}+\ket{11}\bra{00}+\ket{01}\bra{01}-\ket{10}\bra{10}\right), \nonumber\\
\left(\sigma_0-\sigma_1\right)^{\Gamma} &=&
p\left(\ket{01}\bra{10}+\ket{10}\bra{01}+\ket{01}\bra{01}-\ket{10}\bra{10}\right) \nonumber\\
&=& \sqrt{2}p\ket{\xi_0}\bra{\xi_0}-\sqrt{2}p\ket{\xi_1}\bra{\xi_1},
\label{eq:ex31}
\end{eqnarray}
for some orthonormal $\ket{\xi_0}$ and $\ket{\xi_1}$ with
\begin{equation}
\ket{\xi_0}\bra{\xi_0}+\ket{\xi_1}\bra{\xi_1}=\ket{01}\bra{01}+\ket{10}\bra{10}.
\label{eq:ex32}
\end{equation}
Now, let
\begin{eqnarray}
\rho&=&\ket{\phi^+}\bra{\phi^+}\otimes \sigma_0 +\ket{\phi^-}\bra{\phi^-} \otimes \sigma_1
\nonumber\\
&&+\ket{\psi^+}\bra{\psi^+} \otimes \sigma_2+\ket{\psi^-}\bra{\psi^-} \otimes \sigma_3,
\label{eq:ex35}
\end{eqnarray}
where
\begin{eqnarray}
\sigma_2&=&\sqrt{2}p\ket{\xi_0}\bra{\xi_0}+q\ket{00}\bra{00}, \nonumber\\
\sigma_3&=&\sqrt{2}p\ket{\xi_1}\bra{\xi_1}+q\ket{00}\bra{00}.
\label{eq:ex33}
\end{eqnarray}
Then $\|\sigma_{0}-\sigma_{1}\| = 4p > 1/2$, and it follows from Corollary~\ref{Cor:RP} that $K_D(\rho)>0$.

Since
\begin{eqnarray}
\sigma_2-\sigma_3&=&\left(\sigma_0-\sigma_1\right)^{\Gamma}, \nonumber\\
\sigma_2+\sigma_3&=&\sqrt{2}p\left(\ket{\xi_0}\bra{\xi_0}+\ket{\xi_1}\bra{\xi_1}\right)
+2q\ket{00}\bra{00}\nonumber\\
&=&\sqrt{2}p\left(\ket{01}\bra{01}+\ket{10}\bra{10}\right)
+2q\ket{00}\bra{00}\nonumber\\
&=&\left(\sigma_2+\sigma_3\right)^{\Gamma},
\label{eq:ex34}
\end{eqnarray}
$\rho^\Gamma=\rho$, that is, $\rho$ has PPT.
Therefore, $\rho$'s are the PPT states with positive distillable key.

However, since the values $x$, $y$, $z$, and $w$ in Theorem~\ref{Thm:KDW} are
$x=4p$, $y=0$, $z=2\sqrt{2}p+q$, and $w=q$,
\begin{eqnarray}
K_D^{DW}(\sigma_{ABE}^{ccq})&=&1+4p\log_24p+q\log_2q\nonumber\\
&&+(2\sqrt{2}p+q)\log_2(2\sqrt{2}p+q)
\label{eq:ex36}
\end{eqnarray}
is not always positive for $0<q<{(2-\sqrt{2})}/{8}$ as seen in Fig.~\ref{Fig:EX34},
where $\sigma_{ABE}^{ccq}$ is the ccq state for the privacy squeezed state of $\rho$.
\begin{figure}
\includegraphics[angle=-90,scale=0.90,width=\linewidth]{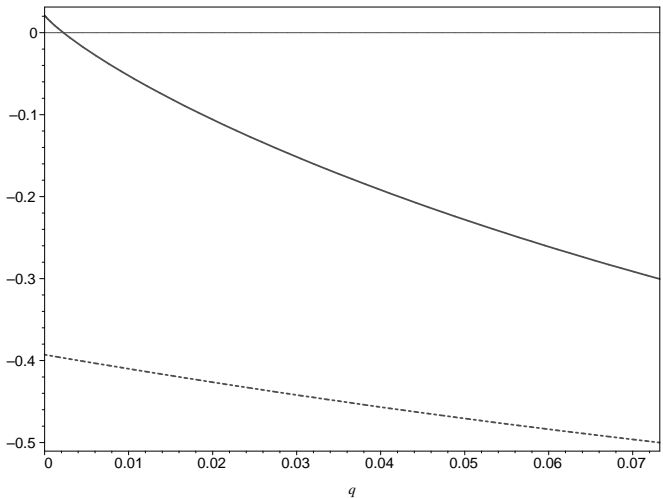}
\caption{\label{Fig:EX34}
The values of $K_D^{DW}$ for the ccq states of the privacy squeezed states:
The solid and dashed curves represent the values of $K_D^{DW}$ for the ccq states
of their privacy squeezed states in Example~\ref{Ex:300} and Example~\ref{Ex:400}, respectively.
}
\end{figure}
Therefore, when $K_D^{DW}<0$, by means of Theorem~\ref{Thm:KDW},
one cannot determine
whether $\rho$ has positive distillable key rate or not,
although one can readily show that $K_D(\rho)>0$ by Corollary~\ref{Cor:RP}.
\end{Exam}

We now present another PPT states with $K_D>0$
but $K_D^{DW}<0$ for the ccq states of their privacy squeezed states.
\begin{Exam}\label{Ex:400}
Let
\begin{equation}
\varrho=\frac{1}{2}\left[%
\begin{array}{cccc}
  {\sigma_0+\sigma_1} & 0 & 0 &  {\sigma_0-\sigma_1} \\
  0 & 2\sigma_2 & 0 & 0 \\
  0 & 0 & 2\sigma_2 & 0 \\
  {\sigma_0-\sigma_1} & 0 & 0 &  {\sigma_0+\sigma_1} \\
\end{array}%
\right],
\label{eq:ex40}
\end{equation}
where
\begin{eqnarray}
\sigma_0&=&p(\ket{\phi^+}\bra{\phi^+}+\ket{01}\bra{01}), \nonumber\\
\sigma_1&=&p(\ket{\phi^-}\bra{\phi^-}+\ket{10}\bra{10}), \nonumber\\
\sigma_2&=&\frac{p}{\sqrt{2}}(\ket{01}\bra{01}+\ket{10}\bra{10}) \nonumber\\
&&+\frac{q}{2}\ket{00}\bra{00}+\frac{q}{2}\ket{11}\bra{11},
\end{eqnarray}
with $p=(1-2q)/(4+2\sqrt{2})$ for $0\leq q<({2-\sqrt{2}})/{8}$.
Then  since
\begin{equation}
\varrho^\Gamma=\frac{1}{2}\left[%
\begin{array}{cccc}
  {\sigma_0^\Gamma+\sigma_1^\Gamma} & 0 & 0 &  0 \\
  0 & 2\sigma_2^\Gamma & {\sigma_0^\Gamma-\sigma_1^\Gamma} & 0 \\
  0 & {\sigma_0^\Gamma-\sigma_1^\Gamma} & 2\sigma_2^\Gamma & 0 \\
  0 & 0 & 0 &  {\sigma_0^\Gamma+\sigma_1^\Gamma} \\
\end{array}%
\right],
\label{eq:ex41}
\end{equation}
and $2\sigma_2^\Gamma\pm(\sigma_0^\Gamma-\sigma_1^\Gamma)$ is positive,
$\varrho$ is a PPT state by Lemma~\ref{Lem:PPT} in Appendix.
Since $\varrho$ satisfies all conditions of Theorem~\ref{Thm:RP},
and therefore $K_D(\varrho)>0$.

However, as in Example~\ref{Ex:300},
we can see that $K_D^{DW}<0$
for the ccq states of their privacy squeezed states.
More precisely, one can obtain that
\begin{eqnarray}
K_D^{DW}&=&1+4p\log_24p\nonumber\\
&&+2(\sqrt{2}p+q)\log_2(\sqrt{2}p+q)
\label{eq:ex42}
\end{eqnarray}
is negative for all $0\leq q<({2-\sqrt{2}})/{8}$, as seen in Fig.~\ref{Fig:EX34}.
\end{Exam}


\section{Summary}\label{Sec:Conclusion}
We have investigated properties of quantum states with positive distillable key rate,
and have constructed sufficient conditions for states to have positive distillable key rate.
Exploiting the conditions,
we have shown that
the PPT bound entangled states given by Horodecki~{\it et al.}~\cite{HHHO1,HHHO2,HPHH}
have nonzero distillable key rate,
and have exhibited a new class of PPT bound entangled states with $K_D>0$,
but with $K_D^{DW}<0$ for the ccq states of their privacy squeezed versions.

\section*{Acknowledgments}
D.P.C. was supported by the Korea Science and Engineering Foundation
(KOSEF) grant funded by the Korea government (MOST) (No.~R01-2006-000-10698-0), and
S.L. was supported by the Korea Research Foundation Grant funded by the Korean Government
(MOEHRD, Basic Research Promotion Fund) (KRF-2006-003-C00044).

\appendix*
\section{Simple Lemmas}
In this appendix, we present some simple but useful lemmas.
\begin{Lem}\label{Lem:PPT}
Let $A$ and $B$ be $n \times n$ hermitian matrices. Then
$
\left[\begin{array}{cc}
A & B\\
B & A\\
\end{array}
\right]
$
is positive if and only if $A \pm B$ is positive.
\end{Lem}

\begin{proof}
Let $\x$ and $\y$ be any vector in $\mathbb{C}^{n}$.
Then we have
\begin{eqnarray}
& &
\left[
\begin{array}{cc}
\x^{\dag} & \y^{\dag}
\end{array}
\right]
\left[%
\begin{array}{cc}
  A & B \\
  B & A \\
\end{array}%
\right]
\left[%
\begin{array}{c}
  \x \\
  \y \\
\end{array}%
\right] \nonumber\\
&=& \x^{\dag}A\x+\y^{\dag}A\y
+\x^{\dag}B\y+\y^{\dag}B\x \nonumber \\
&=& \frac{1}{2}(\x^{\dag}+\y^{\dag})(A+B)(\x+\y)\nonumber\\
&&+\frac{1}{2}(\x^{\dag}-\y^{\dag})(A-B)(\x-\y).
\label{eq:block_matrix_lemma}
\end{eqnarray}
Therefore, we can clearly obtain the proof of this lemma
from Eq.~(\ref{eq:block_matrix_lemma}).
\end{proof}
\begin{Lem}\label{Lem:orthogonal}
For any two positive operators $\sigma_{0}$ and $\sigma_{1}$,
$\|\sigma_{0}-\sigma_{1}\|=\|\sigma_{0}+\sigma_{1}\|$ if and only if
$\T(\sigma_{0}\sigma_{1})=0$.
\end{Lem}

\begin{proof}
If $\T(\sigma_{0}\sigma_{1})=0$, then it is trivial that
$\|\sigma_{0}-\sigma_{1}\|=\|\sigma_{0}+\sigma_{1}\|$.
We now assume that $\|\sigma_{0}-\sigma_{1}\|=\|\sigma_{0}+\sigma_{1}\|$.
Note that
there exist positive operators $\tau_{0}$ and $\tau_{1}$
such that $\sigma_{0}-\sigma_{1}=\tau_{0}-\tau_{1}$
and $\T(\tau_{0}\tau_{1})=0$.
Then from the following equalities
\begin{eqnarray}
\T(\sigma_0)-\T(\sigma_1)
&=&\T(\sigma_0-\sigma_1)=\T(\tau_0-\tau_1)\nonumber\\
&=&\T(\tau_0)-\T(\tau_1),
\label{eq:orthogonal01}
\end{eqnarray}
and
\begin{eqnarray}
\T(\sigma_0)+\T(\sigma_1)
&=&\|\sigma_0+\sigma_1\|=\|\sigma_0-\sigma_1\|=\|\tau_0-\tau_1\|\nonumber\\
&=&\T(\tau_0)+\T(\tau_1),
\label{eq:orthogonal02}
\end{eqnarray}
we obtain $\T(\sigma_0)=\T(\tau_0)$ and $\T(\sigma_1)=\T(\tau_1)$.
Since $\T(\tau_{0}\tau_{1})=0$,
there exist two disjoint index sets $I_0$, $I_1$,
and an orthonomal basis $\left\{\ket{x_j}: {j\in I_0\cup I_1}\right\}$ such that
$\tau_0 = \sum_{j \in I_0} \lambda_j \ket{x_j}\bra{x_j}$ and
$\tau_1 = \sum_{j \in I_1} \lambda_j \ket{x_j}\bra{x_j}$ for some $\lambda_j \geq 0 $.
Then it is straightforwad to have the following equalities.
\begin{eqnarray}
\sum_{j \in I_0}\bra{x_j}\sigma_0\ket{x_j}&-&\sum_{j \in I_0}\bra{x_j}\sigma_1\ket{x_j}
\nonumber\\
&=&\sum_{j \in I_0}\bra{x_j}\sigma_0-\sigma_1 \ket{x_j}=\sum_{j \in I_0}\bra{x_j}\tau_0-\tau_1 \ket{x_j}
\nonumber\\
&=&\sum_{j \in I_0}\bra{x_j}\tau_0\ket{x_j}-\sum_{j \in I_0}\bra{x_j}\tau_1\ket{x_j}
\nonumber\\
&=&\sum_{j \in I_0}\bra{x_j}\tau_0\ket{x_j}=\sum_{j \in I_0\cup I_1}\bra{x_j}\tau_0\ket{x_j}
\nonumber\\
&=&\T(\tau_0)=\T(\sigma_0)
\nonumber\\
&=&\sum_{j \in I_0}\bra{x_j}\sigma_0\ket{x_j}+\sum_{j \in I_1}\bra{x_j}\sigma_0\ket{x_j}.
\label{eq:orthogonal_equalities}
\end{eqnarray}

Hence, we obtain
\begin{equation}
\sum_{j \in I_1}\bra{x_j}\sigma_0\ket{x_j}+\sum_{j \in I_0}\bra{x_j}\sigma_1\ket{x_j}=0.
\label{eq:orhtogonal_final_eq}
\end{equation}
Since $\sigma_0$ and $\sigma_1$ are positive,
it can be obtained that
$\bra{x_j}\sigma_0\ket{x_j}=0$ for any $j \in I_1$, and
$\bra{x_j}\sigma_1\ket{x_j}=0$ for any $j \in I_0$.
Therefore, we conclude that $\T(\sigma_{0}\sigma_{1})=0$.
\end{proof}

Now, we consider the recurrence protocol~\cite{BDSW}
to distill a pbit~\cite{HHHO1,HHHO2}.
By simple but tedious calculations,
we obtain the following lemma.
\begin{Lem}[Recurrence protocol]\label{Lem:RP}
For any  states $\rho_1$ and $\rho_2$ in
$\mathcal{B}(\C^2 \otimes \C^2 \otimes \C^d \otimes \C^{d})$,
let
\begin{equation}
\rho_1
=\left[%
\begin{array}{cccc}
  A_{0000} & A_{0001} & A_{0010} & A_{0011} \\
  A_{0100} & A_{0101} & A_{0110} & A_{0111} \\
  A_{1000} & A_{1001} & A_{1010} & A_{1011} \\
  A_{1100} & A_{1101} & A_{1110} & A_{1111}
\end{array}%
\right],
\label{eq:rho1rho2}
\end{equation}
and
\begin{equation}
\rho_2
=\left[%
\begin{array}{cccc}
  B_{0000} & B_{0001} & B_{0010} & B_{0011} \\
  B_{0100} & B_{0101} & B_{0110} & B_{0111} \\
  B_{1000} & B_{1001} & B_{1010} & B_{1011} \\
  B_{1100} & B_{1101} & B_{1110} & B_{1111}
\end{array}%
\right].
\label{eq:rho2}
\end{equation}
After some LOCC operations in the recurrence protocol,
$\rho_1\otimes\rho_2$ can be transformed into a state in
$\mathcal{B}(\C^2 \otimes \C^2 \otimes \C^{d^2} \otimes \C^{d^2})$.
Let $\rho_{jj}$ be the resulting state when the measurement outcome is $jj$.
Then we have
\begin{widetext}
\begin{eqnarray}
\rho_{00}
&=&\frac{1}{N_0}
\left[%
\begin{array}{cccc}
  A_{0000}\otimes B_{0000}& A_{0001}\otimes B_{0001}& A_{0010}\otimes B_{0010}& A_{0011}\otimes B_{0011}\\
  A_{0100}\otimes B_{0100}& A_{0101}\otimes B_{0101}& A_{0110}\otimes B_{0110}& A_{0111}\otimes B_{0111}\\
  A_{1000}\otimes B_{1000}& A_{1001}\otimes B_{1001}& A_{1010}\otimes B_{1010}& A_{1011}\otimes B_{1011}\\
  A_{1100}\otimes B_{1100}& A_{1101}\otimes B_{1101}& A_{1110}\otimes B_{1110}& A_{1111}\otimes B_{1111}
\end{array}%
\right],
\nonumber\\
\rho_{11}
&=&\dfrac{1}{N_1}
\left[%
\begin{array}{cccc}
  A_{0000}\otimes B_{1111}& A_{0001}\otimes B_{1110}& A_{0010}\otimes B_{1101}& A_{0011}\otimes B_{1100}\\
  A_{0100}\otimes B_{1011}& A_{0101}\otimes B_{1010}& A_{0110}\otimes B_{1001}& A_{0111}\otimes B_{1000}\\
  A_{1000}\otimes B_{0111}& A_{1001}\otimes B_{0110}& A_{1010}\otimes B_{0101}& A_{1011}\otimes B_{0100}\\
  A_{1100}\otimes B_{0011}& A_{1101}\otimes B_{0010}& A_{1110}\otimes B_{0001}& A_{1111}\otimes B_{0000}
\end{array}%
\right],
\label{eq:rho00rho11}
\end{eqnarray}
\end{widetext}
where
\begin{equation}
N_0=\sum_{j,k=0}^1\|A_{jkjk}\|\cdot\|B_{jkjk}\|,
\label{eq:N0}
\end{equation}
and
\begin{equation}
N_1=\sum_{j,k=0}^1\|A_{jkjk}\|\cdot\|B_{\bar{j}\bar{k}\bar{j}\bar{k}}\|,
\label{eq:N1}
\end{equation}
with $\bar{j}=j+1 \pmod 2$.
\end{Lem}


\end{document}